\documentclass[12pt]{iopart}
\usepackage{graphicx}
\usepackage[T1]{url}

\providecommand{\sunmass}{\mathrm{M}_{\scriptscriptstyle\odot}}

\begin{document}
 \title%[Inference on inspiral signals...]
       {Inference on inspiral signals using LISA MLDC data}
 \author{Christian~R\"over$^1$, Alexander~Stroeer$^{2,3}$,
         Ed~Bloomer$^4$, Nelson~Christensen$^5$, 
         James~Clark$^4$, Martin~Hendry$^4$, 
         Chris~Messenger$^4$, Renate~Meyer$^1$, 
         Matt~Pitkin$^4$, Jennifer~Toher$^4$,
         Richard~Umst\"atter$^1$, Alberto~Vecchio$^{2,3}$,
         John~Veitch$^4$ and Graham~Woan$^4$}
 \address{$^1$ Department of Statistics, 
          The University of Auckland,
          Auckland, New Zealand}
 \address{$^2$ School of Physics \& Astronomy, 
          University of Birmingham, Birmingham, UK}
 \address{$^3$ Department of Physics \& Astronomy, 
          Northwestern University, Evanston, IL, USA}
 \address{$^4$ Department of Physics \& Astronomy, 
          University of Glasgow, Glasgow, UK}
 \address{$^5$ Physics \& Astronomy, 
          Carleton College, Northfield, MN, USA}

\begin{abstract}
In this paper we describe a Bayesian inference framework 
for analysis of data obtained by LISA\@.
We set up a model for binary inspiral signals
as defined for the Mock LISA Data Challenge~1.2 (MLDC),
and implemented a Markov chain Monte Carlo (MCMC) algorithm to facilitate
exploration and integration of the posterior distribution over the
9-dimensional parameter space.
Here we present intermediate results showing how,
using this method, information about the 9~parameters 
can be extracted from the data.
\end{abstract}

%  04.80.Nn :  `Gravitational wave detectors and experiments'
%  02.70.Uu :  `Applications of Monte Carlo methods'

\submitto{Classical and Quantum Gravity}

\section{Introduction}
Once the LISA gravitational wave observatory is launched and operational,
it is certain to measure a vast number of signals from a wide range of sources.
Because the data will contain a superposition
of individually modulated signals blended with noise,
sophisticated methods will be required to disentangle individual signals
and consistently infer their parameters.
Bayesian inference provides a means to approach such complex problems,
allowing one to quantify the information that is buried in the data
in a coherent manner~\cite{Jaynes,Gregory,BDA}.
We are convinced that these techniques will be useful, if not essential,
to analyse the data that will be obtained through LISA\@.
Bayesian procedures, in conjunction with Markov chain Monte Carlo (MCMC) methods,
have successfully been applied
for the analysis of ground-based GW 
measurements~\cite{DupuisWoan2005,RoeverMeyerChristensen2006a,RoeverMeyerChristensen2007a},
as well as in the context of
LISA~\cite{WickhamEtAl2006,StroeerGairVecchio2006}, and
in particular in the presence of source confusion~\cite{UmstaetterEtAl2005a}.
Related work on MCMC methods for LISA inspiral analysis can
also be found in \cite{CornishPorter2006b,CornishPorter2007b}.

The authors have gathered as the `\textsl{Global LISA Inference Group}' (GLIG)
and have set out to implement such an analysis framework for LISA data.
We have developed some generic code modules
providing vital components that can be adapted to
allow analysis given different model specifications.
In response to the first round of the Mock LISA Data Challenges
(MLDC) \cite{MLDCoverview}, we present the results of an analysis 
targeted at binary inspiral signals as addressed in MLDC Challenge~1.2.
Within the same context, we approached MLDC Challenge~1.1, 
containing white dwarf binary systems. 
These results are presented in~\cite{StroeerEtAl2007}.

We implemented an MCMC sampler to perform the integration
of the posterior probability distribution of the 9~parameters
that determine the waveform of a binary inspiral GW signal,
and present results illustrating the parameter information 
that can be extracted from the data.
Due to the tight schedule we did not manage
to enhance the MCMC sampler's convergence capabilities sufficiently
to get results for the `blind search' data as well. 
For now we present results for the `training' data only.

\section{Inference framework}
\subsection{The Bayesian approach}
We use a Bayesian framework to perform inference on 
gravitational wave signals observed by LISA,
aiming for the \textsl{information} about parameters
that can be derived from the data.
Information about parameters here is formulated in terms of 
probability distributions over the parameter space.
First the prior knowledge about parameters~$\vartheta$ 
needs to be properly specified
in the \textsl{prior distribution}~$p(\vartheta)$.
Then parameters and data~$y$ are linked by defining 
the \textsl{likelihood}~$p(y|\vartheta)$
that describes how the observables come about for given parameter values.
Inference eventually is done via the parameters' 
\textsl{posterior distribution}~$p(\vartheta|y)$, which
expresses the information about the parameter values 
\textsl{conditional on the oberved data}.
The posterior distribution is given by 
$p(\vartheta|y)\propto p(\vartheta)\,p(y|\vartheta)$, 
as a consequence from \textsl{Bayes' theorem}
\cite{Jaynes,Gregory,BDA}.
Inference on the measured signal's parameters (or other properties)
requires integration of the posterior distribution 
over the parameter space, since one is usually interested in 
determining figures like
posterior expectations, marginal (posterior) densities,
or confidence regions.
We approach the problem using Monte Carlo integration, 
for which we implemented a Markov chain Monte Carlo (MCMC) 
algorithm. 
The algorithm eventually is supposed to
be able to reliably \textsl{find}
the global mode(s) in the posterior distribution
and then perform the \textsl{integration}, 
i.e.\ sample from the posterior
\cite{BDA,McmcInPractice}.

\subsection{Data and parameters}
A gravitational wave (GW) signal is measured by LISA by monitoring
the changes in proper distance between the three satellites as they are orbiting
the Sun. 
The data is sampled every 15~seconds,
which is also about the time it takes for
a photon to travel from one satellite to another.
The measured response is not 
a simple `1:1'~mapping of the signal waveform to the data,
especially when the signal wavelength is of the order or below
LISA's armlength.
Moreover, as LISA orbits, 
the response will also be modulated by Doppler effects 
and be affected by the change in the baseline orientations over time.

The data produced by the spacecraft trio is combined to form
three time-delay-interferometry (TDI) variables, X, Y and Z
\cite{ArmstrongEstabrookTinto1999}.
These can be linearly recombined into three 
stochastically independent components, out of which two 
are sensitive to GW signals (A and E) and one component
is only noise (T)
\cite{PrinceEtAl2002}.
In the following we will only be concerned 
with the former two variables, A and E\@.

In the restricted 2.0~PN approximation,
the 9~parameters defining a binary inspiral's GW signal 
measured by LISA are
chirp mass~($m_c$), mass ratio~($\eta$), coalescence phase~($\phi_c$),
coalescence time~($t_c$), ecliptic latitude~($\theta$), 
ecliptic longitude~($\varphi$), luminosity distance~($D$), 
polarisation~($\psi$) and inclination angle~($\iota$)
\cite{MLDC1doc}.

\subsection{Model}
For given data~$y$ from a single TDI~variable
the likelihood function is defined 
as a function of the parameters~$\vartheta$ by
\begin{equation}
  p(y|\vartheta)  \;\propto\; 
      \exp\left(-\sum_f
                \frac{|\tilde{y}(f)-\tilde{s}(f,\vartheta)|^2}{S_n(f)}\right),
\end{equation}
where $\tilde{y}$ and $\tilde{s}$ are the (numerical) discrete 
Fourier transforms of data and signal waveform respectively, 
and $S_n$ is the variable's (one-sided) noise spectral density \cite{FinnChernoff1993}.
The data~($y$) going into the likelihood here are the `A' and `E' 
TDI~variables \cite{PrinceEtAl2002}.
Assuming that these are stochastically independent, 
the likelihood then is the product of the 
individual variables' likelihoods.
The signal waveform~$\tilde{s}$ to which the data are matched is the 
corresponding TDI response to the GW signal implied by 
the parameters~$\vartheta$.
The noise spectrum~$S_n$ refers to the noise in the corresponding
(A and E) TDI variables.

\section{Implementation}
In order to infer the measured signal's parameters, 
one first needs to \textsl{find} the global mode(s) 
in the posterior distribution, 
and then also to `\textsl{explore}' the mode(s),
i.e.\ simulate posterior samples.
We tackle that problem using MCMC methods, 
an approach that in particular requires
many likelihood evaluations. 
Likelihood computations are computationally expensive, 
since they require several time-consuming steps: 
Given a parameter set~$\vartheta$, one first needs to compute 
the $+$/$\times$ polarisation waveforms emitted by the inspiral event, 
then the TDI~response of the LISA interferometer to the GW signal, 
and finally its Fourier transform, 
before parameters can be related to the data through the likelihood.
Since most of these (and more) steps are common between 
a wide range of different types of analyses,
we set up our software in a modular style so that parts would be
reusable and shareable in form of modules.
See our accompanying paper~\cite{StroeerEtAl2007} for another
application that shares parts of the same code.
So far, this also includes a common framework to store and manipulate data internally,
an interface to the \textsl{lisaXML} data format \cite{MLDC1doc},
and the availability of Fourier transformation and spectrum estimation 
capabilities based on the \textsl{FFTW} library \cite{FFTW}.
Most importantly, the derivation of LISA's response 
(in terms of X/Y/Z or A/E TDI~variables) to a gravitational
wave signal (given in terms of $+$/$\times$ polarisations, 
direction of source, and polarisation angle) was needed.
Here we resorted to 
the \textsl{LISA Simulator} \cite{LISAsimulator211,CornishRubbo2003},
that was also used for the generation of the MLDC data, 
and which is coded in~C, 
allowing us to easily incorporate it into our code and stay
consistent with the provided data.
Originally, the LISA Simulator was not intended to do
its computations repeatedly and quickly, and it was possible to speed up the code
by storing and reusing some intermediate steps. 
\begin{table}[h]
  \caption{\label{tab:speed}Computation speed of the MCMC code 
           for different amounts of data 
           (on an Intel Xeon 2.4$\;$GHz %/ 4$\;$GB 
           processor). 
           Most of the computation time (more than~95\%) goes into deriving 
           A/E TDI responses from the $+$/$\times$ GW waveforms.}
  \begin{indented}
  \item[]\begin{tabular}{cccc} \br
      \multicolumn{2}{c}{amount of data} && seconds \\ \cline{1-2}
      days & samples && per iteration \\ \mr
       364 & $2^{21}$ &&  146 \\
       182 & $2^{20}$ &&   75 \\
        91 & $2^{19}$ &&   38 \\
        46 & $2^{18}$ &&   19 \\
        23 & $2^{17}$ &&   10 \\ \br
    \end{tabular}    
  \end{indented}
\end{table}
We implemented a simple Metropolis-algorithm~\cite{BDA,McmcInPractice} to do inference on 
binary inspiral signals as defined for MLDC challenge~1.2.1 \cite{MLDC1doc}.
The eventual computation speed, depending on how much data are processed,
is shown in table~\ref{tab:speed}.
A similar implementation has proven successful in the context of 
ground-based GW measurements \cite{RoeverMeyerChristensen2007a}, 
and we are currently working on tuning and extending the basic algorithm.

\section{Results}
We applied the above framework to the data in MLDC challenge~1.2.1.
The signal waveform was generated following the description 
given in~\cite{MLDC1doc}, and we estimated the A/E variables' noise spectral 
densities based on the section of data where the signal was absent.

We ran the code on the `training' data set, 
starting from the true parameter values, and, due to the low computation speed,
only considered the last $2^{17}$~samples (corresponding to 23~days of measurements) 
before coalescence for the analysis.
The resulting speed of the MCMC sampler still was rather slow, 
producing a posterior sample every 10~seconds.
By only considering the last part of of the signal before coalescence
we are of course neglecting some information, 
but since the SNR of the injected signal was very high (almost~500),
and most of that is actualised in the last phase immediately before coalescence,
we will still be left with a high SNR\@.
On the other hand, we will especially lose information 
about the location parameters ($\theta$,~$\varphi$), 
since these are encoded in the long-term evolution of the signal,
so we might find an increased degeneracy between these two parameters.
As the efficiency of our code continues to improve
we will of course be analysing larger sections of data.

\begin{figure}[ht]
  \begin{center}
    \includegraphics[width=15cm]{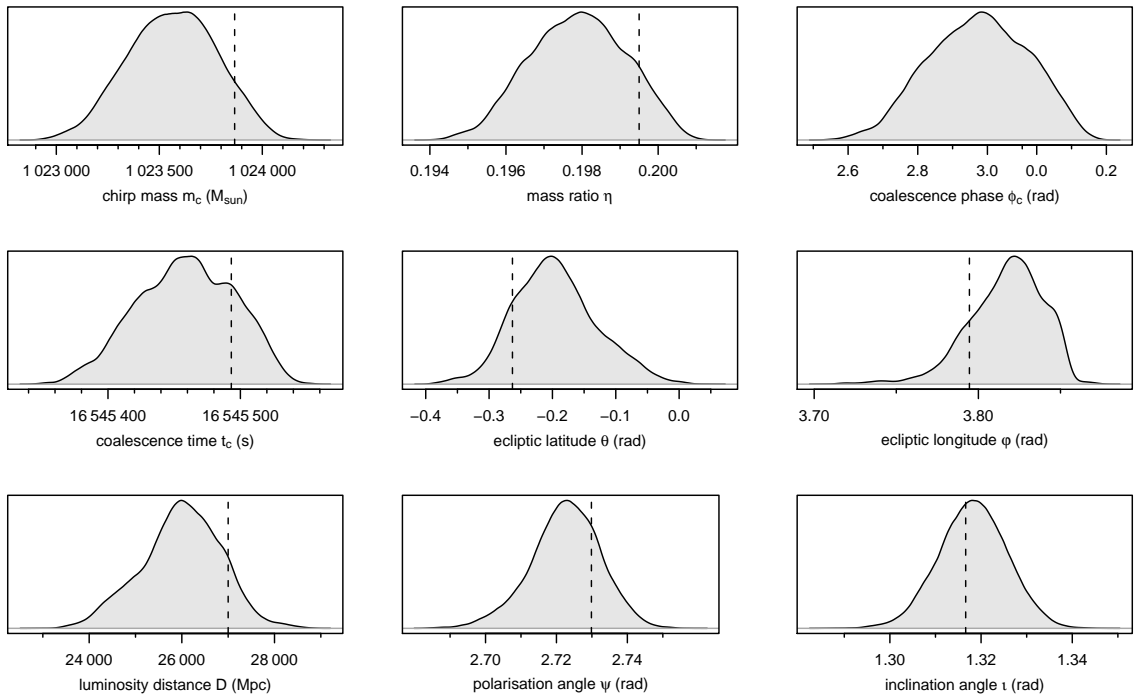}
    \caption{Marginal posterior densities for all 9~parameters.
             Dashed lines indicate the true values.}
    \label{fig:posterior1}
  \end{center}
\end{figure}

Figure~\ref{fig:posterior1} shows the marginal posterior distributions
for all 9~individual parameters in comparison to the true parameter values
(there is no true value shown for~$\phi_c$, since we are using a different 
parametrisation: \textsl{coalescence phase} instead of \textsl{initial phase}). 
\begin{figure}[ht]
  \begin{center}
    \includegraphics[width=15cm]{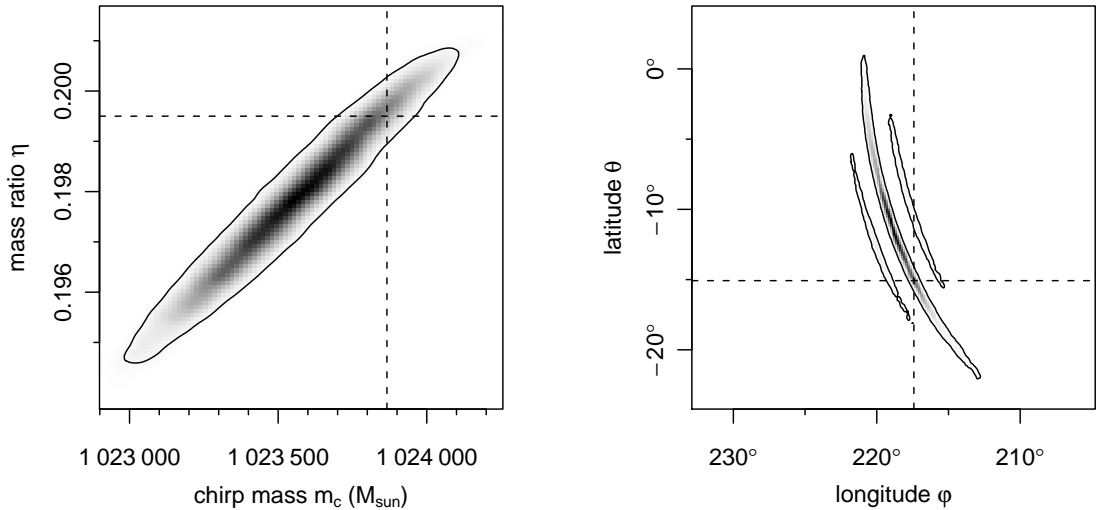}
    \caption{Marginal joint posterior densities 
             and 99\%~credibility regions
             for two pairs of parameters.
             Dashed lines indicate the true values.}
    \label{fig:posterior2}
  \end{center}
\end{figure}
The true parameter values are all well covered by the posterior 
distribution, not only in these 1-dimensional projections, 
but also for all bivariate distributions, 
two examples of which are shown in 
figure~\ref{fig:posterior2};
this demonstrates the consistency of the applied inference framework.
Table~\ref{tab:posterior} shows some summary statistics 
characterizing the posterior distribution, 
and relating it to the true parameter values.
\begin{table}
\caption{\label{tab:posterior}Some summary statistics characterizing 
         the marginal posterior distributions for individual parameters.
         Mean and standard deviation describe location and accuracy,
         and the 99\%~central credibility intervals contain the corresponding
         parameter with 99\%~probability, given the data at hand.}
\begin{center}
\small
\begin{tabular}{rcccccl} \br
                              &  mean          &  st.dev.    &  99\% c.c.i.                     &  true          &  unit \\ \mr
 chirp mass $m_c$             & $1\,023\,564$  & $215$       & ($1\,023\,033$, $1\,024\,054$)   & $1\,023\,866$  & $\sunmass$\\ 
 mass ratio $\eta$            & $0.1979$       & $0.0013$    & ($0.1948$, $0.2006$)             & $0.1995$       & \\
 coalescence phase $\phi_c$   & $2.97$         & $0.14$      & ($2.63$, $0.13$)                 &                & rad\\
 coalescence time $t_c$       & $16\,545\,459$ & $36$        & ($16\,545\,370$, $16\,545\,535$) & $16\,545\,493$ & s\\
 ecliptic latitude $\theta$   & $-0.195$       & $0.065$     & ($-0.357$, $-0.019$)             & $-0.263$       & rad\\
 ecliptic longitude $\varphi$ & $3.817$        & $0.023$     & ($3.736$, $3.859$)               & $3.795$        & rad\\
 luminosity distance $D$      & $25\,991$      & $864$       & ($23\,754$, $28\,195$)           & $27\,000$      & Mpc\\
 polarisation $\psi$          & $2.7226$       & $0.0099$    & ($2.6943$, $2.7464$)             & $2.7299$       & rad\\ 
 inclination angle $\iota$    & $1.3182$       & $0.0075$    & ($1.2978$, $1.3368$)             & $1.3166$       & rad\\
 \br
\end{tabular}
\end{center}
\end{table}
As one can see from figure~\ref{fig:posterior2},
there is much posterior correlation, or degeneracy, 
between the parameters.
In particular, there are two groups of parameters 
that are highly correlated with each other:
firstly the two mass parameters, coalescence time and phase
($m_c$, $\eta$, $t_c$, $\phi_c$),
and secondly, the two sky location parameters and the luminosity distance
($\theta$, $\varphi$, $D$).
Correlation coefficients of parameter pairs within these groups 
are as high as~0.90--0.99, 
which greatly complicates sampling from the posterior.

We also ran the code on the `blind' challenge~1.2.1 data set,
for which we did not know the true parameter values,
but due to the code's speed and the size of the parameter space it would not
converge and produce results in time before the submission deadline
for MLDC round~1.

While the MCMC algorithm is working in principle and performing the
posterior integration, more tuning is necessary to enhance
its optimisation properties, i.e. its capabilities of finding modes by itself, 
and its efficiency in manoeuvering through parameter space.
Better convergence properties are crucial not only to enhance
the algorithm's overall applicability, but also to make sure not to be missing
further posterior modes that may be of relevance.

\section{Conclusions}
We have presented a Bayesian inference framework for analysis of GW signals
as measured by LISA\@.
We ran a basic MCMC algorithm 
on data simulating a binary inspiral measurement
from the first round of the Mock LISA Data Challenges (MLDC).
In a related effort \cite{StroeerEtAl2007}, sharing parts of the same code,
we applied a similar model to the analysis of signals
from white dwarf binary systems.
The MCMC implementation so far is a simple Metropolis algorithm, 
and the results illustrate that this approach ultimately 
allows one to extract and express the information about signal parameters
contained in the data in a coherent manner.
While the \textsl{integration} of the posterior distribution 
over the parameter space is fully functional, 
more work needs to be done on the MCMC sampler's 
\textsl{optimisation} capabilities as well as its efficiency.
We are working on a preprocessing stage to the MCMC algorithm 
to provide rough parameter 
estimates as starting values for the MCMC sampler.
We are also currently extending the Metropolis-sampler
to a \textsl{parallel tempering algorithm}
\cite{McmcInPractice,RoeverMeyerChristensen2007a}
in a parallel implementation
\cite{PaprzyckiStpiczynski}.
The underlying model will also need to be generalised
by including the noise spectrum as an unknown,
which might just mean the introduction of an additional 
`Gibbs step' in the MCMC sampler~\cite{BDA,McmcInPractice}.

\ack
This work was supported by 
the Marsden Fund Council from Government funding 
administered by the Royal Society of New Zealand, 
the National Science Foundation, 
the Fulbright Scholar Program
and the Packard Foundation. 

\section*{References}
  \bibliographystyle{unsrt}
  \bibliography{/home/phd/christian/literature/literature}

\end{document}